\documentstyle[12pt,epsf]{article}

\begin{document}
\begin{center}
{\LARGE \bf Exact results for the universal area distribution of clusters 
in percolation, Ising and Potts models\\}  
\vspace{8mm}
{\large \bf John Cardy\\}
\vspace{2mm}
{Department of Physics\\
Theoretical Physics, 1 Keble Road, Oxford OX1 3NP, UK\\ 
\& All Souls College, Oxford\\
j.cardy1@physics.oxford.ac.uk\\}
\vspace{8mm}

{\bf and}

\vspace{8mm}
{\large \bf  Robert M. Ziff\\}
\vspace{2mm}
{Michigan Center for Theoretical Physics \\
\& Department of Chemical Engineering\\
University of Michigan, Ann Arbor, MI 48109--2136, USA\\
rziff@umich.edu,  734-764-5498 \\} 

\end{center}
\vspace{6mm}
\begin{abstract}
At the critical point in two dimensions, the number of percolation clusters of enclosed area greater than
$A$ is proportional to $A^{-1}$, with a proportionality constant $C$ that is universal.
We show theoretically (based upon Coulomb gas methods), and verify
numerically to high precision, that $C = 1/8\sqrt3\pi
= 0.022972037\dots$. We also derive, and verify to
varying precision, the corresponding constant for Ising spin clusters, and for
Fortuin-Kasteleyn clusters of the $Q=2$, 3 and 4-state Potts models.


Running title:  Cluster-area distribution in percolation, Ising and Potts models.
\end{abstract}

\section{Introduction.}
\label{intro}

It is often useful to characterize critical systems  by
their geometric properties, for example
the distribution of cluster sizes which appears to follow 
a power law
\begin{equation} n_s \sim B s^{-\tau}
\label{ns}
\end{equation}
asymptotically for large $s$, where $n_s$ gives the number of
clusters of $s$ connected sites,  per lattice site.  The exponent
$\tau$ is a universal quantity whose value is the same for all
systems of a given class --- for example, $\frac{187}{91}$
for all critical percolation systems in two dimensions, no
matter what lattice or percolation type is considered as long as the rules
are sufficiently local. The
coefficient or amplitude $B$ however is non-universal, varying
from lattice to lattice.

Indeed, $n_s$ cannot have a completely universal form because it is
written in terms of a lattice-level measure, the mass $s$. Different
lattice structures have different typical site densities at the
lattice level and correspondingly different values of $B$.  In
order to characterize the size distribution of clusters in a way
that circumvents the site-level description, the authors of Ref.\
\cite{ZiffLorenzKleban} considered (for the case of two-dimensional
percolation)
the quantity Nr$(\ell_m > \ell) = \tilde N(\ell)$   which gives the 
number of clusters whose maximum $x$- or $y$- dimension $\ell_m$ is
greater or equal to a given value $\ell$, divided by the total
system area, ${\cal A}=O(L^2)$.  They argued that for $L >> \ell >> a$
(where $a$ is the lattice spacing), this
quantity should behave as
\begin{equation}
\tilde N(\ell) \sim {\tilde C \over \ell^2} \ ,
\label{Nl}
\end{equation}
with the coefficient $\tilde C$ being a universal quantity,
identical for all 2d percolating system at the critical point. The
universality of $\tilde C$ follows heuristically from the idea that
$\tilde N$ represents a  macroscopic measure of the large clusters
of the system, and remains well defined in the limit $a \to 0$,
in which the lattice disappears.  The proportionality to
$1/\ell^2$ is a consequence of the self-similarity of the fractal
percolating system, and can also be derived by the following
argument (in $d$ dimensions): from (\ref{ns}) it follows that the
number of clusters whose mass is greater than
$s$ scales as $s^{1-\tau}$, and because $s \sim \ell^D$,
where $D$ is the fractal dimension of the clusters, the
number of clusters whose length scale is greater than $\ell$ scales
as $\ell^{D(1-\tau)} $, or $\ell^{-d}$ by virtue of the
hyperscaling relation $d/D = \tau - 1$. This result is valid for
any critical system where the hyperscaling relation is valid.
Later we shall give other, presumably equivalent, theoretical arguments.

Besides the maximum dimension $\ell_m$, one can consider any other
macroscopic measure of the length scale of the cluster, such as
the radius of gyration or the diameter of the covering disk.  For each
measure, there is a corresponding value of $\tilde C$.

An equivalent way to write (\ref{Nl}) is
\begin{equation}
N(A) \sim {C \over A}\qquad\quad(a^2\ll A\ll L^2)
\label{Narea}
\end{equation}
where $N(A)$ is the number of clusters (per unit area) whose
area (by some measure) is greater or equal to $A$,
and $C$ depends upon the choice of that measure.
This could be the area of the smallest disk covering the cluster,
the area enclosed by the cluster, and so on.  Eq.\ (\ref{Narea}) is the
form of the size distribution that will be considered
in this paper.

We note that (\ref{Narea}) can also be written as
\cite{ZiffLorenzMandelbrot02}
\begin{equation}
A_n \sim {C \over n}
\label{Zipf}
\end{equation}
for $1 << n << 1/a^2$, where  $A_n$ represents a
rank-ordering of  the areas, such that $A_1$ is the area  of the
largest cluster, $A_2$ the area of the second-largest, etc.,  for a
system whose total area ${\cal A}$ is defined as unity. Although
the rank-ordering necessarily starts with clusters whose area is of
the order of the area of the system, the behavior of (\ref{Zipf})
applies to clusters whose area is much smaller than $1$ but larger
than the lattice element area.  Eq.\ (\ref{Zipf}) gives the size
distribution in a proper Zipf's-law form, in which the weight (here
area) is inversely proportional to the rank. When  written in terms 
of $s$, on the other hand,
the behavior of the size ranking is not a simple inverse
power as above (compare \cite{Watanabe96} and
\cite{JanStaufferAharony98}) and also is not universal.

The various measures of the
area of the clusters that were considered
in Ref.\ \cite{ZiffLorenzMandelbrot02} included the
area of the square $l_m \times l_m$, the
area of a disk that just covers the cluster, the area enclosed by the external
perimeter (hull) of the cluster, and the area enclosed by  the
Grossman-Aharony (G-A) hull of the cluster (in which fjords are excluded)
\cite{GrossmanAharony87}. (Percolation hulls are
fractal with dimension
$D_H = 7/4$ \cite{SaleurDuplantier87,SapovalRossoGouyet85} but
enclose a non-fractal, Euclidean area.)     For each of these
measures, a different value of the constant $C$ applies, and  the
following values were found: $C$(square) $\approx 0.115$, $C$(disk)
$\approx 0.104$, $C$(G-A hull) $\approx 0.037$, and
$C$(hull) $\approx 0.024$.
  The way these different values of $C$ were found was that  the
first $C$(square) was measured directly on a fully populated lattice
(since the measurement of the maximum
$x$- or $y$-direction is an easy task),  and then the rest were
deduced (in an approximate way)
by looking at the ratio of the area measures for
individually generated clusters.  It was noticed that $C$(disk) is
close to the fractal co-dimension $d-D = 5/48 = 0.1041666\ldots$,
but no exact results for any of these $C$ were obtained.

In the present paper we report on a direct numerical and theoretical study of the
constant $C = C$(hull)  for the 2d measure of the area enclosed by
the external perimeter or hull of percolation clusters,
Ising spin clusters, and Fortuin-Kasteleyn (FK) clusters
on the Potts model clusters for $Q = 2$, 3
and 4.  (Of course, percolation corresponds to the Potts model for $Q = 1$ and the
Ising model corresponds to $Q=2$.)  

Initially, one of us predicted that for percolation
\begin{equation}
C = {1\over8\sqrt3\pi} = 0.022972037\ldots 
\label{Cperc}
\end{equation}
Independently, the other
numerically determined
$C = 0.022976 \pm 0.000005$, which is completely consistent with this
prediction.
Additional work described below yields $0.0229723 \pm 0.0000010$ (one
standard deviation of error).  This close agreement confirms that
the Coulomb gas methods that are
used to derive these results are most certainly applicable to
percolation and the Potts model.  We also considered different
lattices and types of percolation to demonstrate universality. 

For Ising clusters of same-spin sites, we predict the value
\begin{equation}
C = {1\over16\sqrt{3} \pi} = 0.011486019\ldots \qquad
{\hbox{(Ising spin clusters)}}
\label{Cising}
\end{equation}
exactly half the value for percolation clusters.
For the Potts model with $Q = 2,3,4$, we
also consider the areas enclosed by the FK bond clusters
\cite{FortuinKasteleyn72}, and find for the corresponding values of $C$:
\begin{eqnarray}
C = {1\over 12 \pi} = 0.026525824\ldots  \qquad &(\hbox{FK cluster, } Q = 2) \\
C ={\sqrt{3}\over 20 \pi} = 0.027566445\ldots  \qquad &(\hbox{FK cluster, } Q = 3) \\
C = {1\over 4 \pi^2} = 0.025330296\ldots \qquad &(\hbox{FK cluster, } Q = 4) \\
\nonumber\end{eqnarray}

The theoretical justifications of the above predictions are based upon
considerations reported previously in \cite{JC2} and expanded
upon in the second section below.
In the third section we describe the numerical work
we carried out to
test these results; we find good numerical confirmation
for all the cases.
Conclusions are given in the fourth section.

\section{Coulomb gas calculation of $C$.}
\label{theory}

In this section we compute the universal amplitude $C$ in the scaling
law $N(A)\sim C/A$, using Coulomb gas methods \cite{Nienhuis}. These are
not rigorous, but are known to give presumably exact results for
critical exponents and other universal quantities.

While the development in the Introduction emphasizes clusters and the
areas enclosed by their external hulls, the focus will shift to both external
and internal hulls, or loops when both are taken together. 
A factor of one-half will be included in the final results
to compensate for this
change, so that the values for $C$
will be applicable to just external hulls,
just internal hulls, or the average (but not sum) of the two.

We consider a finite but large system of linear size $L$, and total
area ${\cal A}=O(L^2)$. As will become
clear, the precise geometry and boundary conditions are not relevant to
the calculation of $C$. $L$ is considered to have dimensions of length,
so that the total number of sites in the lattice is of order $(L/a)^2$,
where $a$ is the lattice spacing.

All the models we consider (percolation, Ising spin clusters, FK
clusters) are special cases of either the O$(n)$ model or the $Q$-state
Potts model \cite{Nienhuis}.

The O$(n)$ model is most easily considered
on a honeycomb lattice, and it is equivalent to the loop gas model
defined by the partition function
\begin{equation}
Z_{O(n)}=\sum_{\rm loop configs} x^{\rm total\ length}\, n^{\rm
number\ of\ loops}
\end{equation}
where the sum is over all configurations of non-intersecting closed
loops on the honeycomb lattice. This model has in general two critical
points for each $n$ in the interval $[-2,2]$:
$x=x_{c1}(n)$ (dilute phase) and $x=x_{c2}(n)$ (dense phase).
In particular, for $n=1$ and $x=x_{c2}=1$ the loops form the hulls of
site
percolation clusters on the triangular lattice; for $n=1$ and
$x=x_{c1}$
they are the boundaries of critical Ising clusters. For $n\to0$ we get
a
single self-avoiding loop, and for $n=2$ the loops are the steps on a
surface at the roughening transition.

The partition function of the $Q$-state Potts model is more easily
considered on the square lattice, and it may be transformed into
that of the random cluster model, proportional to
\begin{equation}
Z_Q=\sum_{\rm cluster\ configurations}x^{\rm number\ of\ bonds}
Q^{\rm number\ of\ clusters}
\end{equation}
where $x=p/(1-p)$. The hulls of the clusters form closed loops on
the medial lattice, a square lattice whose vertices lie at the
midpoint of the links of the original lattice. At the critical
point
$x=x_c(Q)=\surd Q$, $Z_Q$ is proportional to the partition function
of a loop gas
\begin{equation}
Z_Q\propto \sum_{\rm loop\ configurations}(\surd Q)^{\rm number\ of\
loops}
\end{equation}
Note that both internal and external cluster hulls are counted as
loops.

In both cases, then, the critical models are equivalent to loop gases
with fugacity $n$ (resp.~$\surd Q$) per loop. The hulls of the FK
clusters are in the same universality class as the dense phase of the
O$(n)$ model with $n=\surd Q$.

As discussed in Sec.~\ref{intro}, we are interested in the number $N(A)$
of such loops, per unit area of the lattice, whose internal area is
greater than a given $A$. Note that we consider $A$ as having
dimensions (length)$^2$. For $A\ll L^2$, we expect that $N(A)$ has a
finite limit as $L\to\infty$. In order to obtain universal results, we
also consider $A\gg a^2$. Our computation of the form of $N(A)$ in
this regime is in two stages: first we show, from Coulomb gas
arguments, that the total area contained in \em all \em loops behaves
logarithmically, $\propto {\cal A}\ln(L/a)$, as $L/a\to\infty$,
with a calculable
coefficient; then we argue from this that in the regime of interest
$N(A)\sim C/A$, with $C$ simply related to the above coefficient.

\subsection{Total area inside all loops.}
We shall present two \em a priori \em independent arguments, both however
based on Coulomb gas methods, for evaluating the leading behavior of the
the total area in side all loops in  large but finite system.
\subsubsection{Wilson loop method.}
The argument of this section follows that of
Refs.~\cite{JC1,JC2}, but, in order to be self-contained, we
present it again, perhaps with greater clarity.

The loop gases described above may be mapped exactly onto a height
model
on the dual lattice, as follows. Each loop is assigned a random
orientation, so that a configuration of $m$ unoriented loops corresponds
to $2^m$ configurations of oriented loops.
There is a 1-1 mapping between the configurations of this
oriented loop gas and the heights $h$, conventionally chosen to be
integer multiples of $\pi$, as follows: assign $h=0$ on the boundary,
and increase
(decrease) $h$ by $\pi$ each time a loop is crossed which goes to the
left (right). The fact that the loops are closed makes this a
consistent
procedure.

The weights of $n$ (resp.~$\surd Q$) associated with each
loop may be taken into account in the ensemble of oriented loops in (at
least) two ways: the most natural would be to assign equal weights
$\frac12n$ to each orientation: we refer to this as the \em real \em
ensemble, and denote averages with respect to this ensemble with
conventional brackets $\langle\ldots\rangle$. However, these weights have
the considerable disadvantage of not being local when expressed in terms
of the height variables. Instead, for calculational purposes, a different
weighting is usually chosen, in which the phases ${\rm e}^{\pm{\rm
i}\alpha}$ are distributed along each loop
by assigning a phase ${\rm e}^{{\rm
i}\alpha\theta/2\pi}$ each time the loop turns leftwards
through an angle $\theta$.
Each anticlockwise (clockwise) loop thus accumulates a total phase
${\rm e}^{{\rm i}\alpha}$ (resp.~${\rm e}^{-{\rm i}\alpha}$). On
summing
over orientations, these account for the loop weights as long as
\begin{equation}
n=\surd Q=2\cos\alpha
\end{equation}
Clearly, these weights cannot now be interpreted as probabilities, and we
refer to this as the \em complex \em ensemble. Averages with respect to
this will be denoted by $[\ldots]$.

Within each ensemble,
we may view each oriented loop as carrying a unit current in the sense
of its orientation. Let $J_\mu(x,y)$ be the corresponding
current \em density\em.
For example, for a current directed along a link in the positive
$y$-direction, located at $x=0$, $J_x=0$ and $J_y=\delta(x)$. This
current density may be used to give a formula for the area of a single
closed loop:
\begin{equation}
A=-\frac12\int\int|x-x'|\delta(y-y')J_y(x,y)J_y(x',y')dxdydx'dy'
\end{equation}
This formula is valid for any non-self-intersecting loop, and is
independent of its orientation. If however we
now consider the same quantity evaluated for a given configuration of a
gas of many loops, and we sum over the orientation of each loop
independently, $J_y(r)J_y(r')$ will
average to zero if $r$ and $r'$ are on different loops. Thus the
expression
\begin{equation}
\label{intJJ}
-\frac12\int\int|x-x'|\delta(y-y')\langle J_y(x,y)J_y(x',y')
\rangle d^2\!rd^2\!r'
\end{equation}
where $\langle\ldots\rangle$ denotes the average over the loop gas
ensemble,
gives the mean total area $\langle A_{\rm tot}\rangle$ inside \em all
\em
loops.

What is the current density $J_\mu(r)$ in the height model
parametrization of the configurations? Let us imagine that the
definition of the height function $h(r)$ is extended to ${\bf R}^2$
in such a way that it is constant within each plaquette.  An
obvious candidate is then $J_\mu\equiv
(1/\pi)\epsilon_{\mu\nu}\partial_\nu h$. 
Within the real ensemble, this is clearly correct. It is easy to see that
$\langle J_\mu(r)\rangle=0$ on summing over orientations of a given loop
which passes through $r$.
However, in general $[J_\mu(r)]\not=0$. Consider the average over
orientations in the complex ensemble for a \em fixed \em configuration of
unoriented loops:
\begin{equation}
[J]\propto 1\cdot{\rm e}^{{\rm i}\alpha}+
(-1)\cdot{\rm e}^{-{\rm i}\alpha}\not=0
\end{equation}
Instead, we have to consider a different operator as representing the
current in the complex ensemble:
$\tilde j_\mu\propto\epsilon_{\mu\nu}\partial_\nu{\rm e}^{-2{\rm
i}\alpha h/\pi}$, which now gives
\begin{equation}
\label{tildej}
[j]\propto \big({\rm e}^{-2{\rm i}\alpha}-1\big){\rm e}^{{\rm
i}\alpha}
+\big({\rm e}^{2{\rm i}\alpha}-1\big){\rm e}^{-{\rm i}\alpha}
=0
\end{equation}
as required. In Coulomb gas language, $j$ has charge
$-2{\rm i}\alpha/\pi$. Since there must be overall charge
neutrality,  this is balanced by a charge $+2{\rm i}\alpha/\pi$
distributed on the boundary.

However, the orientation-averaged
two-point function $\langle J(r)J(r')\rangle$ is \em not \em correctly represented
by $[j(r) j(r')]$, since once again this has the wrong charge.
Instead we should take
\begin{equation}
\langle J_\mu(r)J_\nu(r')\rangle=\lambda[J_\mu(r) j_\nu(r')]
\label{twopoint}
\end{equation}
Note that this does vanish when $r$ and $r'$ are on different loops, by
virtue of (\ref{tildej}).
The constant $\lambda$ is fixed by requiring that, for a fixed long loop
whose
sides at $x$ and $x'$ are parallel to the $y$-axis, after summing over
orientations,
$\langle J_y(x,y)J_y(x',y)\rangle=-n\delta(x)\delta(x')$.
The factor $n$ arises from the sum over loop orientations in the height
model.
This gives
\begin{equation}
\lambda\left(\big(1-{\rm e}^{-2{\rm i}\alpha}\big){\rm e}^{{\rm
i}\alpha}
-\big(1-{\rm e}^{2{\rm i}\alpha}\big){\rm e}^{-{\rm i}\alpha}\right)
=-n
\end{equation}
so that $\lambda=n/(4{\rm i}\sin\alpha)$.

All of this is exact, on the lattice. The weights in the height model
are
local but complicated, involving as they do the phase factors ${\rm
e}^{\pm{\rm i}\alpha}$. The central assumption of the Coulomb gas
approach is that, for the purposes of studying the long-distance
behavior of correlation functions, they may be replaced by the
continuum measure
$\exp(-S)$, with $S=(g/4\pi)\int(\partial h)^2d^2\!r$. The parameter
$g$ may be determined by a number of methods \cite{Nienhuis,Kondev} to
give $g=1-\alpha/\pi$, so that $n=\surd Q=-2\cos(\pi g)$. The correct
branches are $1\leq g\leq2$ for $x_{c1}$, and $0\leq g\leq1$ for
$x_{c2}$.

Within this free field theory, it is straightforward to compute
the correlation function $[J_\mu(r)j_\nu(r')]$
in terms of the Green function
$G(r-r')=[h(r)h(r')]\sim-(1/g)\ln|r-r'|$.
First note that
\begin{eqnarray}
[h(r){\rm e}^{-2{\rm i}\alpha h(r')/\pi}]
&=&\sum_{p=0}^\infty{(-2{\rm i}\alpha/\pi)^p\over p!}
[h(r)h(r')^p]\\
&=&\sum_{p=0}^\infty{(-2{\rm i}\alpha/\pi)^p\over p!}
pG(r-r')[h(r')^{p-1}]\\
&\sim&(2{\rm i}\alpha/\pi g)\ln|r-r'|[{\rm e}^{-2{\rm i}\alpha
h(r')/\pi}]\\
&=&(2{\rm i}\alpha/\pi g)\ln|r-r'|
\end{eqnarray}
The last equality follows because of the way the phase factors enter
the
sum over orientations, so that $[{\rm e}^{-2{\rm i}\alpha
h(r')/\pi}]=1$.
We thus find that
\begin{eqnarray}
[J_\mu(r) j_\nu(r')]
&=& (2{\rm i}\lambda\alpha/\pi^2
g)\epsilon_{\mu\kappa}\epsilon_{\nu\sigma}
\partial_\kappa\partial'_\sigma\ln|r-r'|\\
&=& (2{\rm i}\lambda\alpha/\pi^2
g)\epsilon_{\mu\kappa}\epsilon_{\nu\sigma}
\left({2R_\kappa R_\sigma\over R^4}-{\delta_{\kappa\sigma}\over R^2}
\right)
\end{eqnarray}
where we have introduced $R=r-r'$. After a little algebra we therefore
find
\begin{equation}
\label{JJk}
\langle J_\mu(r)J_\nu(r')\rangle
=k(n){R_\mu R_\nu-\frac12R^2\delta_{\mu\nu}\over R^4}
\end{equation}
This form of the
2-point correlation function of a conserved current is in fact
dictated by rotational invariance, but it is the
coefficient $k(n)$ which is the main result:
written in terms of $g$ it is\footnote{There is a misprint in the
corresponding equation in Ref.~\cite{JC2}.}
\begin{equation}
\label{kn}
k(n)={n(1-g)\over \pi g\sin(\pi g)}={2(g-1)\over \pi g}\cot(\pi g)
\end{equation}

Substituting this result into (\ref{intJJ}), we notice that the result
would appear to diverge logarithmically at $r=r'$. However, this is
an artifact of the continuum approximation: the result (\ref{JJk})
is valid only for separations $|r-r'|\gg a$. Since the potential
divergence is logarithmic, the amplitude of the leading term is
insensitive to the precise nature of the modification at shorter
distances, and therefore we may impose a simple cut-off $|r-r'|>a$ on
the
integral. Similarly, the precise form (\ref{JJk}) becomes invalid
for separations $O(L)$, but the short-distance leading logarithm $\ln
a$
 must always appear in the form $\ln(a/L)$, on dimensional grounds,
independent of the precise geometry. Thus the mean total area
within all loops behaves as
\begin{equation}
\label{Atot}
\langle A_{\rm tot}\rangle=(k(n)/2){\cal A}\ln(L/a)+O(1)
\end{equation}
where $\cal A$ is the total area of the system.

As shown in Ref.~\cite{JC2}, in any simply connected region $\cal R$
the
right-hand side of (\ref{Atot}) is proportional to
$\sum_m(1/\lambda_m)$,
where the $\lambda_m$ are the eigenvalues of $-{\rm laplacian}$ in
$\cal R$, with Dirichlet boundary conditions. The leading term always
has the universal logarithmic behavior shown above. Up to a
non-universal constant which may be absorbed into the cut-off $a$,
the $O(1)$ remainder is universal and depends only on the shape of
$\cal R$. For example, for a rectangle it is related to modular forms.
\subsubsection{Relation between total area and the mean depth.}
\label{depth}
In this section we show how the leading behavior of $\langle A_{\rm
tot}\rangle$ may also be found using methods of conformal field theory on
a cylinder, as an extension of the results of Ref.~\cite{Cardy00}.
Let us define, for a given configuration of unoriented loops, the
`depth' $d(r)$ of a given site on the dual lattice to be the minimum
number of loops which must be crossed to connect $r$ to the boundary. That
is, it is the number of noncontractible loops surrounding $r$. In the
height description, it is the supremum of $h(r)/\pi$ over all possible
orientations of the given set of loops. 
As with $h(r)$, we may extend the domain of definition of $d(r)$ 
to the continuum plane by assuming that it is constant over each plaquette
of the dual lattice. Then a little thought shows that the total area
within all loops is simply
\begin{equation}
A_{\rm tot}=\int d(r)d^2\!r
\end{equation}
so that we need to evaluate $\langle d(r)\rangle$.  We do this by first
evaluating $\langle d(r,r')\rangle$, where $d(r,r')$ is the minimal number
of loops which separate distinct points $r$ and $r'$ in the infinite
plane. By translational invariance this is a function of $r-r'$ only.
Let us conformally map the plane into a cylinder of perimeter $2\pi$
via the usual mapping
$w=\ln z$. As $|r-r'|\to\infty$ the images of these points are far apart
along the cylinder, and $d(r,r')$ is therefore asymptotically the same
as the number of loops which wrap around the cylinder between these
points. 

>From the point of view of the height model with complex weights, these
loops must in any case be treated separately, since the factors of 
${\rm e}^{\pm{\rm i}\theta\alpha/2\pi}$ all cancel, so that each orientation is,
a priori, counted with weight 1. As is well known, this may be
compensated for by inserting operators ${\rm e}^{\pm{\rm i}\alpha h/\pi}$
at opposite ends of the cylinder. The free energy per unit length then
is $-c/12$, where $c=1-(6/g)(\alpha/\pi)^2$. The first term comes from the
Casimir effect of the fluctuations of the $h$-field, while the second is
the correction due to the flux between the charges at either end. This
then gives the correct result for the central charge $c$. 

Let us now count the loops which wind around the cylinder with a weight
$n'=2\cos\alpha'$, instead of weight $n=\surd Q=2\cos\alpha$. The free
energy per unit length will be simply modified by an additional term
\begin{equation}
\delta f=(1/2\pi^2g)({\alpha'}^2-\alpha^2)
\end{equation}
The mean number of loops which wrap around the cylinder, per unit length,
is then found by taking $n'(\partial/\partial n')$ of this expression
and setting
$n'=n$. Transforming back to the plane, $\langle d(r,r')\rangle$ is given
by this same coefficient, 
multiplying $\ln(|r-r'|/a)$. After a little algebra
we then find
\begin{equation}
\langle d(r,r')\rangle\sim \big(k(n)/2\big)\ln(|r-r'|/a),
\end{equation}
where $k(n)$ is given by Eq.~(\ref{kn}).
The logarithmic dependence of this function also follows 
from the work of Ref.\ \cite{ChayesChayesDurrett}, who 
derived that dependence from general scaling arguments
but did not find the coefficient given above.

This result, of course, applies to the number of loops separating $r$
and $r'$ in the infinite plane. However, notice that it has a
logarithmically divergent dependence on $a$, which comes from loops
which are much smaller in size than $|r-r'|$. This same divergence
should arise if we now consider the number of loops separating a given
point $r$ from the boundary in a large but finite system, for all points
whose distance from the boundary is much larger than $a$ (but still much
less than $L$). We conclude that 
\begin{equation}
\int\langle d(r)\rangle d^2\!r\sim-\big(k(n)/2\big){\cal A}\ln a
\sim \big(k(n)/2\big){\cal A}\ln(L/a),
\end{equation}
where the last statement holds because $L$ is the only dimensionful
parameter available to compensate $a$. This gives a second derivation of
Eq.~(\ref{Atot}).

We note in passing that $\langle d(r,r')\rangle=
(1/\pi)^2\langle\big(h(r)-h(r')\big)^2\rangle$, in the height
representation. The fact that this behaves logarithmically with $|r-r'|$
is consistent with the hypothesis that, in the continuum limit, the
heights are distributed according to a gaussian ensemble 
$\exp\big(-(1/2\pi k(n))\int(\nabla h)^2d^2\!r\big)$ in the \em real \em
ensemble, even though the lattice weights are nonlocal. However, this
hypothesis is incorrect: as may be shown by extending the above
calculation on the cylinder to higher moments, the cumulant 
\begin{equation}
\langle\big(h(r)-h(r')\big)^4\rangle-
3\langle\big(h(r)-h(r')\big)^2\rangle^2\sim {\rm const.\ }
\ln(|r-r'|/a),
\end{equation}
and does not vanish as it would in a gaussian ensemble.
However, note that this cumulant decays
faster than each term on the left hand side, so that asymptotically
the distribution of $d(r)$ is normal, as was proved by Kesten and Zhang
\cite{KestenZhang}.

\subsection{Relation between $k$ and $C$.}

In this section we first show how the logarithmic behavior of
$\langle A_{\rm tot}\rangle$ provides further justification for the
assertion that $N(A)\sim A^{-1}$ for $a^2\ll A\ll L^2$, then show
how to relate the coefficients.
Recall that $N(A)$ is the number of loops with area greater than
$A$, divided by the total area of the system. For $A\sim L^2$, this
will
also depend on $L$, so let us write it as $N(A,L)$. On dimensional
grounds it has the form
\begin{equation}
N(A,L)=(1/A)F(A/a^2,L/a)
\end{equation}
where $a$ is the lattice spacing.
For $a^2\ll A\ll L^2$, we expect it to be independent of $L$, but, a
priori, it could depend on $a$. In this regime,
let us suppose it has the form
\begin{equation}
\label{form}
N(A,L)\sim 2C(1/A)(A/a^2)^\omega
\end{equation}
where $C$ is a constant and $\omega$ is some exponent.
This is not of course the most general dependence which is possible,
but
is that which would arise if, for some reason, the area scaled
non-trivially with $a$, that is, had a fractal structure. Such
dependence, with $\omega\not=0$, would for example occur in the
distribution of masses, rather than of areas, of percolation clusters.
The form (\ref{form}) should of course connect smoothly onto the
behavior for $A\sim a^2$, when we expect that $N\to {\rm const.}$,
and $A\sim L^2$, where $N\to 0$.

Now the total area $A_{\rm tot}$ within all loops is related to $N$
by
\begin{equation}
A_{\rm tot}=\sum_AN(A,L)
\end{equation}
Comparing with (\ref{Atot}), we see that the contribution to the sum
from the region $a^2\ll A\sim L^2$ will exceed $O(\ln(L/a)$ if
$\omega>0$, and similarly the contribution from
$a^2\sim A\ll L^2$ will violate this bound if $\omega<0$.
Therefore $\omega=0$, and $N(A)\sim 2C/A$ for $a^2\ll A\ll L^2$.
Admittedly, this argument assumes the \em ansatz \em (\ref{form}),
and the reader may be more comfortable with the hyperscaling argument
put forward in the Introduction. However, independently of the
validity of (\ref{form}), our argument shows that if $N(A)\sim C/A$,
then the coefficient $C$ is related to $k(n)$. For then 
the leading contribution from the region $a^2\ll A\ll L^2$ is 
$2C\ln({\cal A}/a^2)\sim 4C\ln(L/a)$, so that comparing once again with
(\ref{Atot}),
\begin{equation}
C=k(n)/8\quad,
\end{equation}
with $k(n)$ given by (\ref{kn}).

For percolation cluster hulls and FK clusters in the $Q$-state Potts model,
we take $n=\surd Q = -2\cos \pi g$ in the dense phase $0\leq g\leq 1$,
which yields $g=\frac23$, $\frac34$, $\frac56$, $1$ for $Q = 1,$ 2, 3, and 4,
respectively.
For critical Ising spin clusters, we take $n = 1$ in the dilute phase
where $1\leq g\leq 2$, so that
$g=\frac43$.  Then by (\ref{kn}) we find the values of $C$ given in
Sec.\ \ref{intro}  and also listed in Table \ref{tableC} (taking the limit
in the case $Q = 4$).  The logarithmic corrections that appear for
the case $Q = 4$ are derived in the Appendix.

\begin{table}
\begin{center}
\begin{tabular}{|c|c|c|}
\hline
cluster type & C(theoretical) & C(measured) \\ \hline
Percolation & ${1/8\sqrt{3}\pi} = 0.022972037\ldots$ & $ 0.0229721(1) $\\
Ising spin & ${1/16\sqrt{3}\pi} = 0.011486019\ldots$ & $ 0.01149(5) $\\
Ising FK & ${1 / 12 \pi} = 0.026525824\ldots  $ & $ 0.0265 $\\
$Q=3$ Potts FK & ${\sqrt{3}/ 20 \pi} = 0.027566445\ldots $ & $ 0.0278 $\\
$Q=4$ Potts FK & ${1/ 4\pi^2}= 0.025330296\ldots $ & $ 0.0258 $\\
\hline
\end{tabular}
\end{center}
 \protect\caption[2]{\label{tableC}
{Predicted and measured values of $C$ for various systems}.}
\end{table}

\section{Numerical results.}
\label{numerical}

To test these predictions, we carried out numerical studies of
percolation on square and triangular lattices with both site and
bond percolation, and the Ising/Potts models
on the square lattice.
For percolation we considered two
ways to generate the clusters: populating the entire lattice,
and individual hull generation.

\begin{figure}
\centerline{
\epsfxsize=4in
\epsfbox{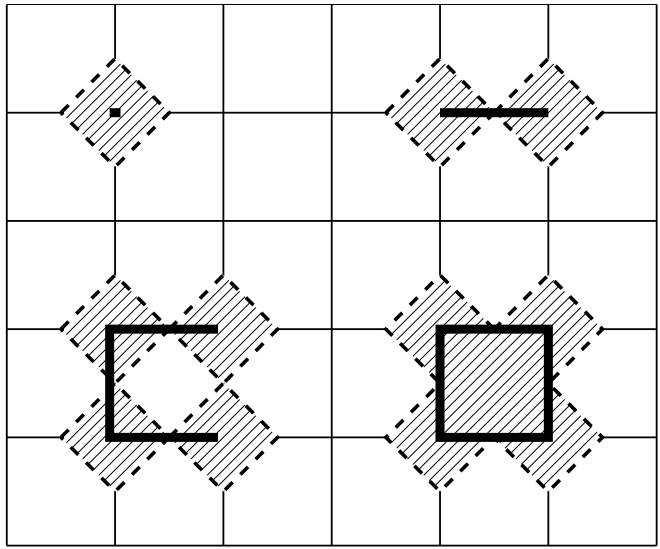}}
\caption{Hull paths for bond percolation, with enclosed  shaded
areas of $\frac12$ (top left), 1 (top right), 2 (bottom left)
and $2\frac12$ (bottom right).  These are all external hulls --
the last case also has an internal hull of area $\frac12$ (not shown).}
\label{figbondlat}
\end{figure}

\subsection{Bond percolation --- Full-lattice population method.}

In the full-lattice population method, we first assign all bonds on
the lattice as occupied or vacant with probabilities $p$ or $1-p$,
respectively, and then carry out all possible hull walks around these bonds.
These walks go from the center to center of
each bond along the diagonals, as shown in Fig.\ \ref{figbondlat},
and turn by an
angle $+\pi/2$ when the center of an occupied bond is encountered,
and by
$-\pi/2$ when the center of  a vacant bond is encountered.
Each walk is completed when it returns to its beginning step.

For
bond percolation on the square lattice, we used a square lattice of
size $512 \times 512$ with periodic boundary conditions, with $p$ at
the threshold $1/2$.
We simulated $10^7$ samples, amounting to a total of
$5.2\times10^{12}$ bonds occupied or not.
We used the R9689 random number generator of Ref.\
\cite{Ziff98}. In the computer program we employed an array of size
$2048 \times 2048$, so that we had distinct array locations to
represent the bonds and each diagonal leg of the hull walks.

The enclosed area of a hull walk was found ``on the fly"
by the following
method: Initially, the area is set equal to zero. When the walk steps to
the right, the area is increased by one-half the $y$ coordinate at
the center of the diagonal step (where we had an array point), and
decreased by one-half the $y$ coordinate when the walk steps
left. The zero point of the $y$ coordinate is irrelevant, because
its value cancels out.  The factor of $1/2$ comes from the fact
that each leg of the hull walk changes the $x$-coordinate by $\pm
1/2$; we are taking the spacing of the bond lattice to be unity.
When the walk closes, this algorithm gives the area of the enclosed
space, with a sign attached:  positive areas correspond to external
hulls that surround clusters, and negative areas corresponds to
internal hulls (which are of course external to the clusters on the
dual lattice).

The smallest area is 1/2; for positive area this corresponds to the
hull around an isolated site (one with no bonds attached), and
negative 1/2 corresponds to the hull inside a square of four
occupied bonds, or equivalently around an isolated site on the dual
lattice. The area of all the hulls are in units of 1/2.
(Alternately, one could consider the lattice spacing to be $\surd{2}$;
then the hulls would all have integer areas, and the system area 
would be $2L^2$.) 

Because we use periodic boundary conditions, there is the
possibility that some hulls could wrap around the torus once or
more before closing into a loop.  The areas for such loops are
undefined, unless taken in pairs, but in any case
we discarded them because we
are interested in clusters whose size is much smaller than the
size of the system.

We found the statistics for internal and external  hulls were
identical (within numerical error), as one would expect for this
self-dual system, and  took the average of the two.

For small $A$ we kept track of the quantity $N_A =$ the number of
loops (per unit area) whose enclosed area is exactly $A$,
where $A = {1\over2}, 1, {3\over2}, 2, ...$.
According to (\ref{Narea}), this quantity should behave as
\begin{equation}
N_A = N(A) - N(A + 1/2) \sim {C \over 2A^2}
\label{NA}
\end{equation}
so that $2A^2 N_A \sim C$ for large $A$.
The results are given in Table \ref{perctable1} for $A \le 5$.

\begin{table}
\begin{tabular}{l|l}
$A$ & $N_A(p)$ \\
\hline
 $\frac12$ & $q^4$ \\
  1 & $2 p q^6$ \\
 $\frac32 $& $ 6 p^2 q^8$ \\
 2 & $p^3 (4 q^9 + 18 q^{10}) $ \\
 $\frac52 $& $ p^4 (q^8 + 32q^9 + 55 q^{10}) $ \\
 3 & $p^5 (8 q^{10} + 30q^{12} + 160 q^{13} + 174 q^{14})$ \\
 $\frac72 $& $ p^6 (12 q^{11} + 40q^{12} + 332 q^{14} + 672 q^{15} + 570 q^{16}) $ \\
 4 & $ 2 p^6 q^{11} + p^7 (2 q^{10} + 136q^{13} + 168 q^{14}
+ 336 q^{15} + 2030 q^{16} + 2712 q^{17}$\\
& \hspace{9.2mm}$+ 1908 q^{18})$ \\
 $\frac92 $& $ 20 p^7 q^{13} + p^8 (22 q^{12} + 186q^{14} + 844 q^{15}
+ 868 q^{16} + 4064 q^{17} + 9972 q^{18} $\\
& \hspace{9.2mm}$+ 10880 q^{19} + 6473 q^{20}) $ \\
\end{tabular}
\caption{Exact results for $N_A(p)$ for bond percolation on the square
lattice at occupancy $p = 1 - q$, for $A=\frac12\ldots\frac92$.}
\label{exactpolys}
\end{table}

\begin{figure}
\centerline{
\epsfxsize=2in
\epsfbox{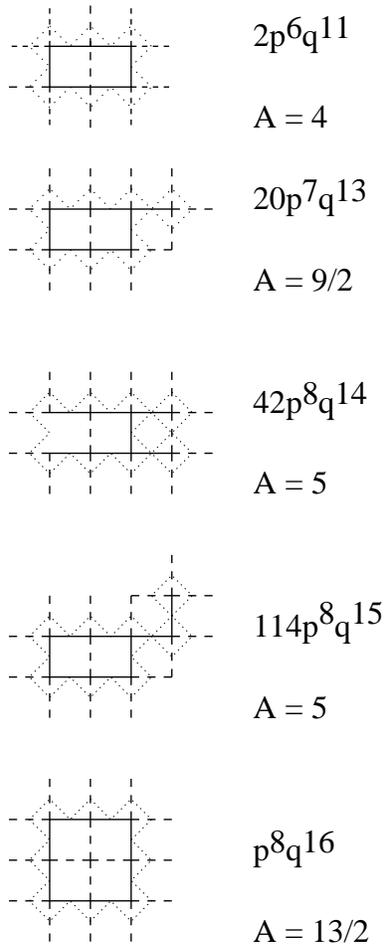}}
\caption{Clusters contributing higher-area terms to polynomials
in Table \protect\ref{exactpolys}.  Solid lines represent
occupied bonds, dashed lines are vacant bonds, and the dotted
lines trace out the external hull.  These are the graphs that
have to be ``moved'' in the usual cluster polynomials
from $A = (b+1)/2$ (where $b$ is the
number of bonds) to higher $A$ due to the existence of enclosed
open spaces. }
\label{graphfig}
\end{figure}

To check these results, we derived the  exact expressions for
$N_A$ for $\frac12 \leq A \leq \frac92$ given in Table \ref{exactpolys}.
These are for an arbitrary bond occupancy of $p$, with $q = 1-p$.
For $A = {1\over2} \ldots 3$ these expressions are identical
to the expressions for the number
of clusters (per site) containing $b = 2A-1$ bonds, which
are well known \cite{SykesGauntGlen,NewmanJensenZiff}.
For larger $A$ we had to make
modifications to the bond cluster expressions to take into account
graphs that contain internal open spaces with vacant bonds,
which result in areas larger than $(b+1)/2$.
We subtracted
the term $2 p^6 q^{11}$ from $N_{7\over2}$ and added it to $N_4$ to
account for the area of an open $1 \times 2$ rectangle,
whose external hull area is 4, not $7/2$.
Likewise, the term $20p^7 q^{13}$ (the $1 \times 2$ rectangle
with an extra bond attached) was subtracted from $N_{4}$ and
added to $N_{9/2}$.  Finally, the terms $42p^8q^{14}$, $114p^8q^{15}$, and
 $p^8q^{16}$, which correspond to various graphs with area
greater than $9/2$, were subtracted from
$N_{9/2}$.  These various diagrams are shown in Fig.\ \ref{graphfig}.
 This shifting of terms has the effect of
making $N_A$ follow asymptotically
the exponent $-2$ of (\ref{NA}) rather than
the exponent $-\tau = -2.055\ldots$ followed by $n_s$.

Taking $p=1/2$ and multiplying by $2A^2$, we arrive at
the estimates for $C$ listed in Table \ref{perctable1}.
The agreement with our numerical results
is excellent --- within the small statistical error.
Interestingly, the convergence of these estimates is
rather quick --- already, at $A = 5$, the result is
within 6\% of the (presumably) exact value.

To analyze the data for larger $A$, we considered the quantity
$N(A,2A) \equiv $ the number of clusters whose enclosed area is greater or
equal to $A$
and less than $2A$.
According to (\ref{Narea}), this quantity should behave as
\begin{equation}
N(A,2A) = N(A) - N(2A) \sim {C \over 2A}
\label{NA2A}
\end{equation}
so that $2 A N(A,2A) \sim C$ for large $A$.

\begin{figure}
\centerline{
\epsfxsize=4.5in
\epsfbox{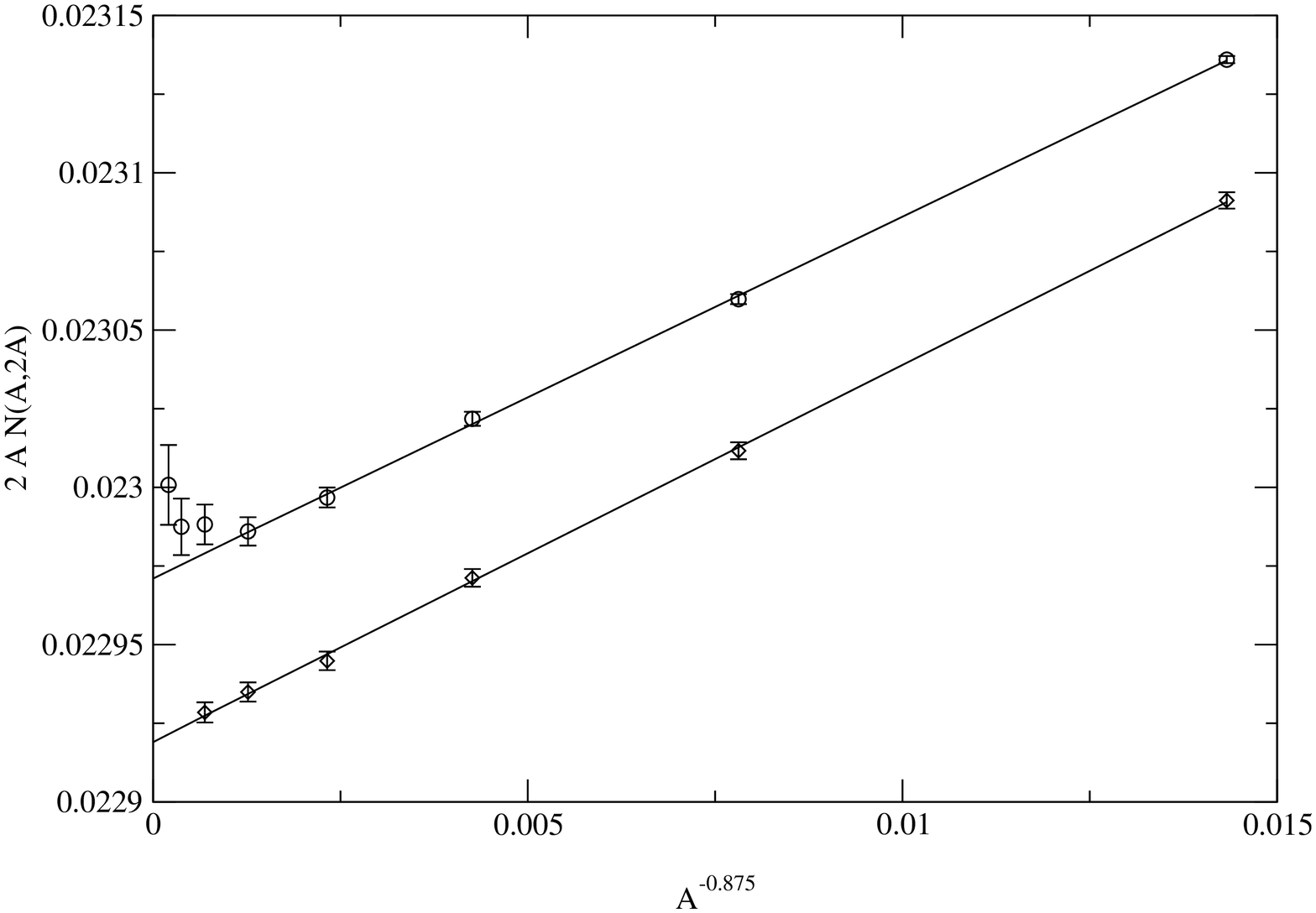}}
\caption{Plot of $2 A N(A,2A)$ vs.\ $A^{-0.875}$
for bond percolation on a square lattice.  Upper data
points: lattice population method.  Lower points (shifted down by 0.00005):
single hull generation method.}
\label{figBSQboth}
\end{figure}

The measured values of $2 A N(A,2A)$ are given in
Table \ref{perctable2}. They  monotonically decrease
to a value 0.0229860(45) for $A = 2048$, but then slightly
increase at $A = 4096$; for larger $A$, the
increase continues, as seen in Fig.\ \ref{figBSQboth},
(upper curve)
where the data from $A = 128$ to 16384
are shown.  We attribute this increase to
interference  of clusters with themselves around the periodic boundary
conditions, and thus ignore these data.
Fitting the  the 5 data points from
$A = 128$ to $2048$ as a function of $A^{-\theta}$ to
a straight line,
we find a good linear fit with $\theta  = 0.875$ as shown in
that figure, with the equation of the
line given by
\begin{equation}
2A N(A,2A) = 0.0229712 + 0.01148A^{-0.875}
\label{bondperclatfig}
\end{equation}
implying $C = 0.0229712$.

 We estimate
the error in the above value of $C$ to be $\approx 10^{-6}$
from the statistical error of the data and the uncertainty in
the extrapolation to infinity. The predicted value
(\ref{Cperc}) falls within these error bars.

\begin{table}
\begin{center}
\vspace{3mm}
\begin{tabular}{|c|c|c|c|}\hline
$A$ & Full & Single & Exact \\
& lattice & hull & results \\ \hline
$ 1/2	$&$	0.0312500	(1)	$&$	0.031247	(3)	$&$	 1/32 = 	0.03125	$\\
$ 1	  $&$	0.0312500	(1) $&$	0.031252	(4)	$&$	 1/32 = 	0.03125	$\\
$ 3/2	$&$	0.0263674	(1)	$&$	0.026363	(4)	$&$	 27/1024 = 	0.026367188	$\\
$2	   $&$	0.0253907	(2)	$&$	0.025398	(5)	$&$	13/512 = 	0.025390625	$\\
$ 5/2 $&$	0.0257491	(2)	$&$	0.025744	(6)	$&$	3375/2^{17} = 	0.025749207	$\\
$3	   $&$	0.0254749	(3)	$&$	0.025477	(6)	$&$	3339/2^{17} = 	0.025474548	$\\
$ 7/2	$&$	0.0249188	(3)	$&$	0.024917	(7)	$&$	104517/2^{22} = 	0.024918795	$\\
$4	   $&$	0.0249898	(4)	$&$	0.025004	(7)	$&$	6551/2^{18} = 	0.024990082	$\\
$ 9/2	$&$	0.0247714	(4)	$&$	0.024778	(8)	$&$	13298985/2^{29} = 	0.02477129	$\\
$5	   $&$	0.0245659	(5)	$&$	0.024557	(8)	$&$				$\\
\hline
\end{tabular}
\end{center}
 \protect\caption[2]{\label{perctable1}
Values of $2 A^2 N_A$ for small $A$ for bond percolation on the square
lattice: two algorithms and exact results.  Errors in last digit are given in
parentheses.}
\end{table}

In terms of the length
scale $\ell \sim A^{\frac12}$, this exponent corresponds to a
correction of the order $\ell^{-1.75}$, which is the
scaling of the hull of the cluster.  Indeed, this finite-size
correction can be interpreted as a surface effect  \cite{Huber},
reflecting the arbitrariness in locating where precisely the hull
of the cluster should be placed.

As a test of our procedure, we also compared  our
measurement of the total number of loops (hulls of both type)
 per unit
area ($\approx$ twice the number of clusters) with the
theoretical result,  which for bond percolation on
the square lattice is given by the Temperley-Lieb
result
\cite{TemperleyLieb,ZiffFinchAdamchik}
\begin{equation}
\sum_A N_A \sim {3 \surd{3} - 5 } =
0.196152422\ldots
\label{TL}
\end{equation}
Our measured value was $0.1961572(14)$, larger 
than the above prediction by only
$0.0000048(28)$.  This difference
corresponds to an  excess number of 1.2 loops per
lattice (found by multiplying the latter number by $512^2$),
which is barely discernible
above the statistical error of $\pm 0.7$.  
In fact, this correction can also be predicted theoretically.
For a system with a rectangular boundary of aspect
ratio $r$, the excess number
of clusters is a known function $b(r)$
\cite{ZiffFinchAdamchik,KlebanZiff}.  To find the excess
number of loops, note that the quantity $n_c + n_{c'} - n_l$
(the number of clusters, plus the number of dual lattice clusters,
minus the number of loops) equals 1 if there is a cross-configuration
on the lattice or the dual lattice, and zero otherwise.
Thus it follows that the excess number of loops is just 
$2 b(r) -2 \pi_+(r)$, where $\pi_+(r)$ is the cross-configuration
probability, which has been calculated by Pinson \cite{Pinson}.
For a square system, the excess number of loops
is predicted to be
\begin{equation}
2 [b(1) - \pi_+(1)] = 2 (0.883576 - 0.309526) = 1.14810
\end{equation}
using $b(1)$ from \cite{KlebanZiff} and $\pi_+(1)$ from  \cite{ZiffLorenzKleban}.
This prediction 
 happens to coincide almost exactly
with the measured value (even though
the error bars of the latter are quite large).
This predicted value can be tested to higher
precision most easily
by going to smaller lattices.

Besides the problem of clusters interfering with themselves,
there is also the problem in the population method
that the statistics for larger hulls are rather poor
because of the relatively small number of such hulls
that are generated.
In the next section we consider a method that addresses
both of these problems.

\begin{table}
\begin{center}
\vspace{3cm}
\begin{tabular}{|c|c|c|}\hline
$A$	&	Full			&	Single					\\
	&	lattice			&	hull					\\		\hline
$ 1/2	$&$	0.0625001	(1)		$&$	0.0625022		(	42	)	$\\
$1	$&$	0.0429689	(1)		$&$	0.0429634		(	32	)	$\\
$2	$&$	0.0306645	(2)		$&$	0.0306677		(	26	)	$\\
$4	$&$	0.0270220	(2)		$&$	0.0270226		(	25	)	$\\
$8	$&$	0.0250527	(3)		$&$	0.0250528		(	25	)	$\\
$16	$&$	0.0240647	(4)		$&$	0.0240635		(	25	)	$\\
$32	$&$	0.0235504	(6)		$&$	0.0235463		(	26	)	$\\
$64	$&$	0.0232785	(8)		$&$	0.0232740		(	27	)	$\\
$128	$&$	0.0231360	(11)		$&$	0.0231412		(	28	)	$\\
$256	$&$	0.0230598	(16)		$&$	0.0230616		(	29	)	$\\
$512	$&$	0.0230218	(22)		$&$	0.0230212		(	31	)	$\\
$1024	$&$	0.0229968	(32)		$&$	0.0229948		(	32	)	$\\
$2048	$&$	0.0229860	(45)		$&$	0.0229849		(	33	)	$\\
$4096	$&$	0.0229882	(63)		$&$	0.0229785		(	35	)	$\\
\hline
\end{tabular}
\end{center}
 \protect\caption[2]{\label{perctable2}
 Values of $2 A N(A,2A)$ for bond percolation on the square
lattice for the two algorithms.}
\end{table}

\subsection{Bond percolation --- Single hull generation method.}

It is well known that percolation clusters can be grown
individually through a process where bonds are made occupied
or not only when they are encountered (the ``Leath" method).
In  the same way,
percolation hulls can be generated individually on a
blank (undetermined) lattice by a kind of growing self-avoiding
walk that mimics the walk used to trace out hulls
\cite{ZiffCummingsStell84}.  For critical
bond percolation
 \cite{Grassberger86}, the walker moves along the edges
of a square lattice (the diagonals in Fig.~\ref{figbondlat}),
and turns by $+\pi/2$ or $-\pi/2$ randomly at each vertex, except
at vertices previously visited, where it always turns
to avoid retracing itself.   The walk terminates when
it returns to the origin
and cannot proceed further.
Note that pseudo-random numbers are generated
only for the bonds that are visited during the walk,
making this method efficient.
This walk has also been studied as a kinetic
Lorentz-gas model \cite{RuijgrokCohen},
and the results here apply to that model also.

In order for the contribution of a given hull to be the same
as on the fully populated lattice, it is necessary to
weight each walk by $1/t$, where $t$ is the number of hull steps.
This compensates for the fact that a hull of $t$ steps
is generated with $t$ times the probability in the single
cluster method compared to the population method, because
there are $t$ places a given walk can start from.

This weighting can also be checked as follows:
The probability of generating a closed hull of
at least $t$ steps is generated with a probability \cite{Ziff86}
\begin{equation}
P(t) \sim  c t^{-1/7}
\label{Pt}
\end{equation}
where $c$ is a constant.
Defining a Euclidean length
scale $\ell \sim t^{1/D_H}$, it follows that the area enclosed
by the walk scales as $A \sim \ell^2 \sim t^{2/D_H} = t^{8/7}$.
Thus, the probability of growing a walk
enclosing at least area $A$ scales as $A^{-1/8}$,
so that the probability
of growing a walk of exactly area $A$ scales as $A^{-9/8}$.
When we weight a hull by the factor
$1/t \sim A^{-7/8}$, we thus get the
proper probability $A^{-9/8-7/8} = A^{-2}$ as given in (\ref{NA}).

In our simulations, we considered square lattices of size $L \times
L$ with periodic b.c.  This is the lattice of the hull walks, which
is rotated by $\pi/4$ from the square bond lattice, and has a spacing
that is $\sqrt{2}/2$ of the bond lattice spacing.
Note that the square system boundary here corresponds to a diamond
on the bond lattice.
We stopped all walks that did not
close by 65536 steps, and kept track of the areas of
all the walks that closed before this cutoff, without
wrapping around the
periodic b.c.
With this cutoff, we could be assured that
all walks that were stopped at the cutoff would
ultimately enclose
an area of at least
$65536/8 = 8192$, taking into account that
there are at most 4 hull steps around each wetted
site, and
each square on the hull-walk (rotated) lattice
 corresponds to an area of $1/2$.

While the statistics of walks of areas smaller than $A = 8192$
should thus be unbiased by having this cutoff, they
can still be biased by the finite-size of the lattice.
For runs on lattices of size $1024 \times 1024$ and smaller,
we found both wraparound clusters and large deviations in
the hull statistics for larger $A$.  Even for lattices of size
$2048\times2048$, where no wraparound occurred with this
cutoff, we still found significant, obviously finite-size
deviations even for $N(A,2A)$ for $A$ below
$A = 8192$.  We attribute these deviations to hulls making contact
with themselves around the periodic b.c., without
actually closing to wrap around.
Therefore, to be absolutely certain of no finite-size
effects, we went to a lattice of size $65536\times65536$
using the virtual lattice method of \cite{ZiffCummingsStell84}.
We checked that with the cutoff of 65536 steps, indeed no walk
got anywhere near the boundary of the system.

We carried out $1.8 \times 10^9$ walks on this lattice,
which, like the simulations for the
$10^7$ fully populated lattices,
took several weeks of workstation computer time.
A total of $3.2 \times 10^{13}$ hull steps simulated
here, compared with $1.0 \times 10^{13}$
in the simulations of the populated lattices.
The algorithm for the single-hull method is
somewhat simpler and more efficient than that for
the lattice population method.

In the  single-hull method, larger hulls
are generated with a higher
probability than in the lattice-population method:
the number generated in the interval $(A,2A)$ (before reweighting)
is proportional to
$A^{-1/8}$ here, compared with $A^{-1}$.
This is advantageous because the large hulls with
their small finite-size effects are essential for finding
$C$ accurately.  On the other hand, in the
single-hull method
 a large fraction
of time is spent on the walks that reach the cutoff
before they close (and are discarded):
Eq.\ (\ref{Pt}) implies that
the total number of steps for all the hulls that reach the cutoff $t_{\rm max}$
grows as 
 $\sim c\, t_{\rm max}^{6/7}$, while the total number of steps for all the hulls that
close before $t_{\rm max}$ is given by
\begin{equation}
\int^{t_{\rm max}} t \left(-{dP \over dt}\right) dt \sim {c \over 6}\, t_{\rm max}^{6/7} \ .
\end{equation}
Thus, no matter what the value of the cutoff is, a fraction $6/7 = 85.7$ \% of the work
(ignoring finite-size effects)
is  spent generating walks that reach the cutoff without closing
and are thus discarded.
Still, for very large cutoffs this overhead is compensated by the
increase in useful statistics for large $A$, making this method
advantageous.

Note that, in our simulations of $3.2\times 10^{13}$ hull-walk steps,
the fraction of those steps belonging to clusters that reached
the cutoff $t_{\rm max} =
65536$ was $6.000124/7$, with the deviation from $6$ in the numerator
being about equal to the apparent statistical error,
$\approx0.0001$. This result seems to provide a very precise
confirmation that the exponent in $P(t)$ is indeed $-1/7$ (i.e., the
hull fractal dimension is $D_H = 7/4$), although to quantify the
precision of this result one would have to investigate different
values of the cutoff $t_{\rm max}$ to determine the finite-size
corrections.

For small $A$, results for $2 A^2 N_A$ are given in Table \ref{perctable1} agree
with the exact values, confirming that the $1/t$ weighting is
correct. Because the single hull method gives fewer of these
small  hulls than the lattice population method, these results have
larger error bars. Here use used $(N_A(\rm total))^{-1/2}$, where
${N_A(\rm total)}$ is the total number of clusters of size $A$,
to estimate the error bars.

Likewise, the results for $N(A,2A)$
for all $A$, given in Table \ref{perctable2}, are seen to agree with
the lattice population results.  For the largest size ranges, the
single-hull method is seen to give better error bars (and are
not biased by finite-size boundary effects).

\begin{figure}
\centerline{
\epsfxsize=3in
\epsfbox{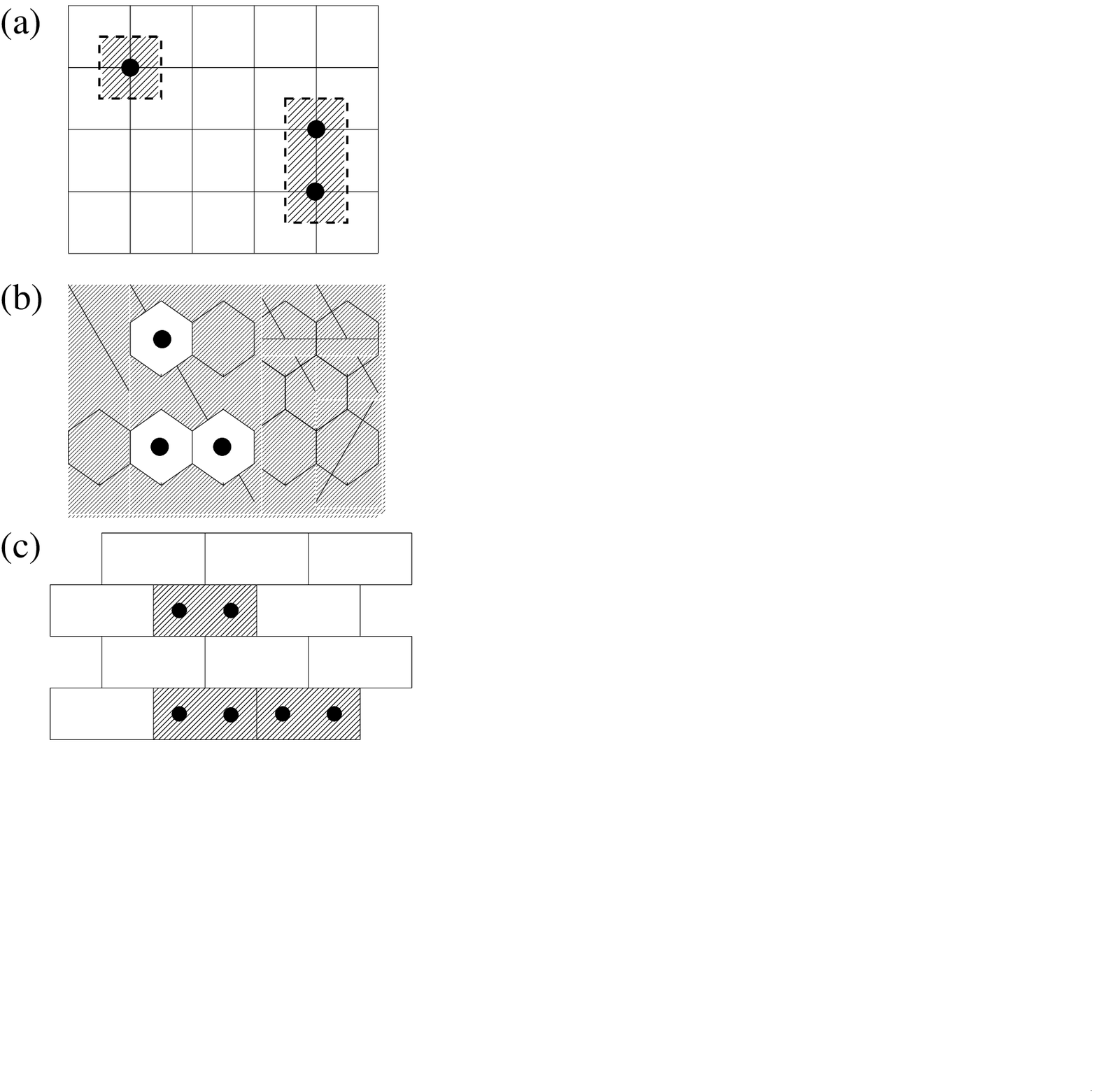}}
\caption{Medial lattices used for hulls in site percolation:
(a) square lattice, (b) triangular lattice, and (c) brick-lattice
form of triangular lattice.}
\label{figsiteperclat}
\end{figure}

A plot of the $2AN(A,2A)$  vs.\ $A^{-0.875}$ for $128 \le A \le 4096$
is also given in Fig.~\ref{figbondlat} (shifted down by 0.00005),
and the data are fit by the linear function
given y
\begin{equation}
2A N(A,2A) =  0.0229692 + 0.01197 A^{-0.875}
\end{equation}
which is consistent with the results of the lattice population method
(\ref{bondperclatfig}).  The error bars on the intercept is about the same,
$10^{-6}$.

Thus, although the single-hull method is in principle advantageous,
for the system size we considered we obtained $C$ with about the
same precision as the lattice population method, with about the
same amount of work.  However, the single-hull method allowed us
to show that the curvature in the behavior of $2 A N(A,2A)$
for large $A$  as seen in Fig.\ \ref{figbondlat} was indeed due
to cluster interference around the periodic boundaries.

Assuming the predicted value of $C$ given by (\ref{Cperc}),
we can also make a plot of $\log(2 A N(A,2A) - C)$ vs.\ $\log(A)$
(not shown);
with single-hull data we find good linear behavior with a slope
$-\theta = -0.88 \pm 0.01$, which is consistent with the value $0.875$ that
we have been using.

\subsection{Site percolation on the square and triangular lattices.}

We also carried out simulations of site percolation on
two different lattices to demonstrate the universality of the result
(\ref{Cperc}) for $C$.

For site percolation, the logical choice for the hull walk around a
cluster is to follow  a path on the medial lattice whose vertices
are at the center of the faces of the lattice, as shown
in Fig.\ \ref{figsiteperclat} for  the square and triangular lattice.  This choice
allows the single  isolated site to have a non-zero area, and is
symmetric for internal and external hulls for the triangular
lattice.

For the square lattice, we carried out $4\times10^8$ samples on a
lattice of size $256 \times 256$, using the weighted single-hull
method. (In our program we employed a computer array of size $512 \times 512$ to
include the sites of the  medial lattice.)   With such a small
lattice, finite-size effects  appeared for hulls with $A$ larger than
$\approx 1024$.  We used occupancy probability
$p = 0.592746$, which is close to the critical threshold for this
system \cite{NewmanZiff}.   Here we generated the hulls starting
from a segment between a single occupied and vacant site,
which occurs in a populated system  with a probability of
$p(1-p)$.  The latter  factor was therefore
included in the total weight of each hull, along with the
$1/t$ weight, where here $t$ is the number of steps along
the medial lattice.

We found that the statistics for internal and external hulls are
quite different, as one would expect by the asymmetry of this system.
For example, for $A = 1$, $N_1 = p(1-p)^4\approx 0.0163053$ for
an external hull, and $N_1 = (1-p)p^8\approx 0.00620604$ for an internal
hull. This large difference persists as $A$ increases, and suggests
that some other definition  of the hull which gives more symmetric results
between external and internal hulls might be advantageous.

\begin{figure}
\centerline{
\epsfxsize=4.5in
\epsfbox{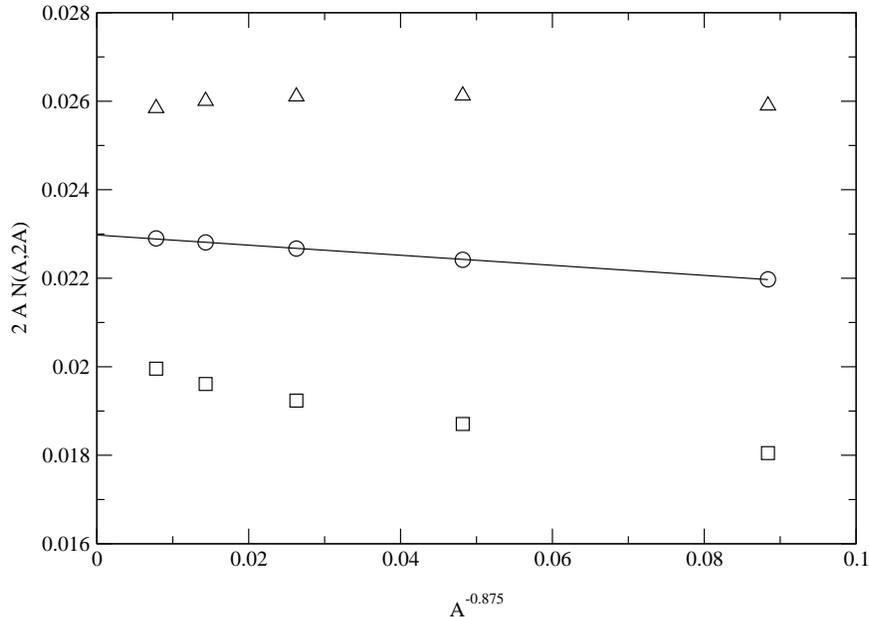}}
\caption{Plot of $2 A N(A,2A)$ vs.\ $A^{-0.875}$ for site percolation
clusters on a square lattice:
external hulls (triangles), average (circles) and internal hulls (squares).
The equation of the line fit through the average points is
given in Eq.\ (\ref{ssqfit}).}
\label{figSSQ}
\end{figure}

In Fig.\ \ref{figSSQ} we show
$2AN(A,2A)$ for the two kinds of hulls,  along with their average.
Taking the average
is the same as including both types of hulls in the area
calculation (and dividing by two).  Indeed, in the  theoretical
development in Section \ref{theory}, both internal and external
hulls were included in the calculation, so it is
appropriate to take this average.   The finite-size corrections to the average
measure again followed a behavior with exponent close to $-0.875$,
which was used in the plot in Fig.\ \ref{figSSQ}.  The line in that
figure is fit by the equation
\begin{equation}
2A N(A,2A) = 0.022976 -0.0114 A^{-0.875}
\label{ssqfit}
\end{equation}
where the unit of area is one square lattice spacing on the
square lattice.
The average measure extrapolates
(for large $A$) to a value $\approx 0.022976$,  in  obvious agreement with the
theoretical prediction, and making a more precise determination
rather superfluous.

Note that, the coefficient to the correction term is similar
in value to the
coefficient for bond percolation, even though a different kind of path
was used to define the hulls in the two cases.  The similarity might be a
coincidence, or it might reflect a fundamental equivalence of perimeter
corrections for site and bond percolation on this lattice.

For site percolation on the triangular lattice the
medial lattice is
a  honeycomb lattice with a hexagon around each vertex of the
triangular lattice as shown in Fig.\ \ref{figsiteperclat}. To
implement this in the computer, we used the square-lattice form
of the honeycomb lattice, also shown in that figure, where
the hexagons become rectangular bricks and a single site on the
triangular lattice now becomes  a pair of sites on the square
lattice.  Thus,
 we could use the same basic algorithm as we used
 for site percolation on the  square lattice, with the only
modification being that sites are occupied or made vacant in
pairs, with a probability 1/2.

For this case (the triangular lattice) we used the lattice-population
method on an underlying square lattice  of size $1024 \times 1024$
with periodic b.c.   We generated $2.4 \times 10^6$ independent
samples. As expected, the internal and external hulls had equal
statistics, within error, reflecting the symmetry of this system.

Again, the data closely followed the $A^{-0.875}$ behavior,
and we do not plot it.  Fitting the data in the range $2^6 < A
< 2^{12}$ (where $A$ is measured in square-lattice units, so that the
smallest hexagon corresponds to $A=2$) we found the following behavior:
\begin{equation}
2A N(A,2A) = 0.022977 -0.0146 A^{-0.875}
\end{equation}
again agreeing with the predicted value of $C$. In this case,
that value is approached from below for finite systems,
with the definition of cluster-hull area used here.

\subsection{Ising clusters.}

To study the clusters of the Ising model, we considered a square
lattice of size $1024 \times 1024$ with periodic b.c., and
simulated the system at the critical temperature of $\exp(-J/kT) = 1 +
\surd{2}$ (where $J$ is the coupling constant in the Potts model
formulation, $H = -J\sum \delta_{\sigma_i}\delta_{\sigma_j}$)
 using the Wolff variation \cite{Wolff} of the Swendsen-Wang method
\cite{SwendsenWang}.  We initialized the lattice with  1000 updates,
and then measured the hull area distribution treating the
system exactly as if it were one of site percolation, using the same
definition of hull areas as shown in Fig.~2 and indeed the same
algorithm. This was followed by 10--100  Wolff updates, and the
 procedure was repeated.  140,000 realizations were generated.

\begin{figure}
\centerline{
\epsfxsize=4.5in
\epsfbox{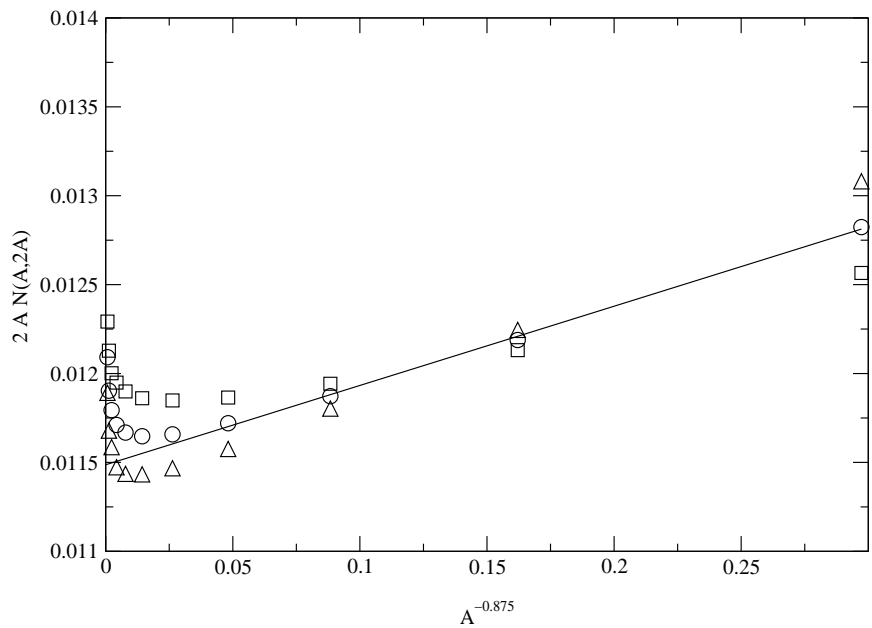}}
\caption{Plot of $2 A N(A,2A)$ vs.\ $A^{-0.875}$ for Ising clusters:
external hulls (triangles), average (circles) and internal hulls (squares).
 The line is fit through the four rightmost points of the average values.}
\label{figisingC}
\end{figure}

As in the site percolation case, we found rather large differences
between internal and external hulls, as seen
in Fig.~\ref{figisingC}.  Here we also found very large deviations
for large $A$, presumably reflecting stronger correlations due to
the interaction.  (Indeed, runs on a smaller $256 \times 256$ lattice
showed even stronger large-$A$ deviations.)
The average measure for the smaller hulls is consistent with
the 
$A^{-0.875}$ finite-size scaling used in that figure, and a
fit of the points for $A = 2$ through 16
yields the
straight line as shown in that figure, with a fit of
\begin{equation}
2A N(A,2A) = 0.011487 + 0.004458 A^{-0.875}
\end{equation}
The intercept is nearly identical with the predicted value of $C$
(\ref{Cising}), in spite of the rather small size of clusters
that were used.  We estimate
the error to be $\pm 0.00005$.

\subsection{FK Clusters on the Potts model for $Q = 2$, $3$ and $4$.}

We also studied the FK clusters on the Potts model at the
critical temperature $e^{J/kT} = 1 + \surd{Q}$.   These clusters
are the bond percolation clusters when bonds are drawn between
neighboring identical spins with a probability $1 - e^{-J/kT} =
\surd{Q}/(1 + \surd{Q})$. We  defined the hulls exactly as in
the square-lattice bond percolation case (Fig.~1) and indeed could use the same
algorithm to trace out and measure the hulls after the bonds
have been specified.

\begin{figure}
\centerline{
\epsfxsize=4.5in
\epsfbox{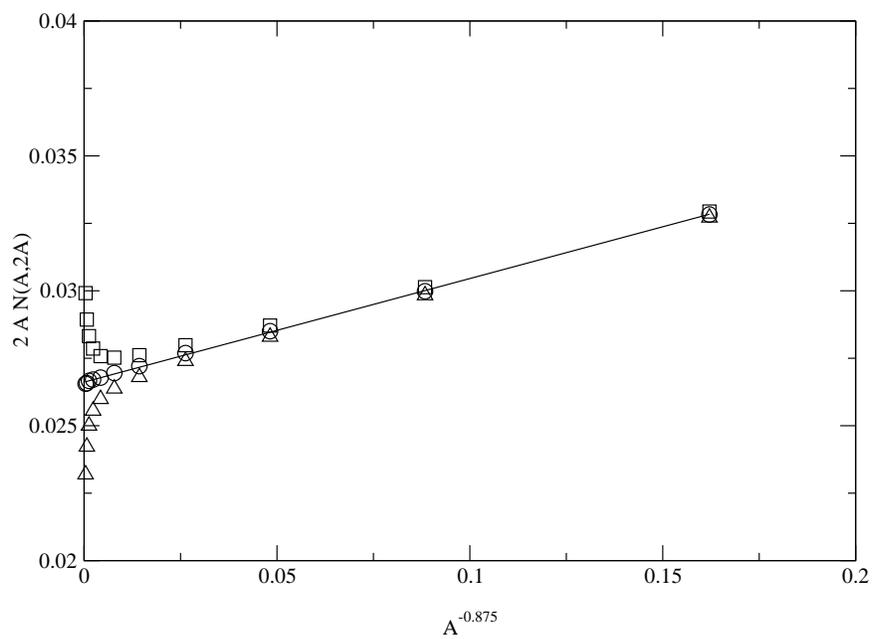}}
\caption{Plot of $2 A N(A,2A)$ vs.\ $A^{-0.875}$ for FK clusters of
the Ising model (Q=2 Potts):
external hulls (triangles), average (circles) and internal hulls (squares).}
\label{figFKising}
\end{figure}

\begin{figure}
\centerline{
\epsfxsize=4.5in
\epsfbox{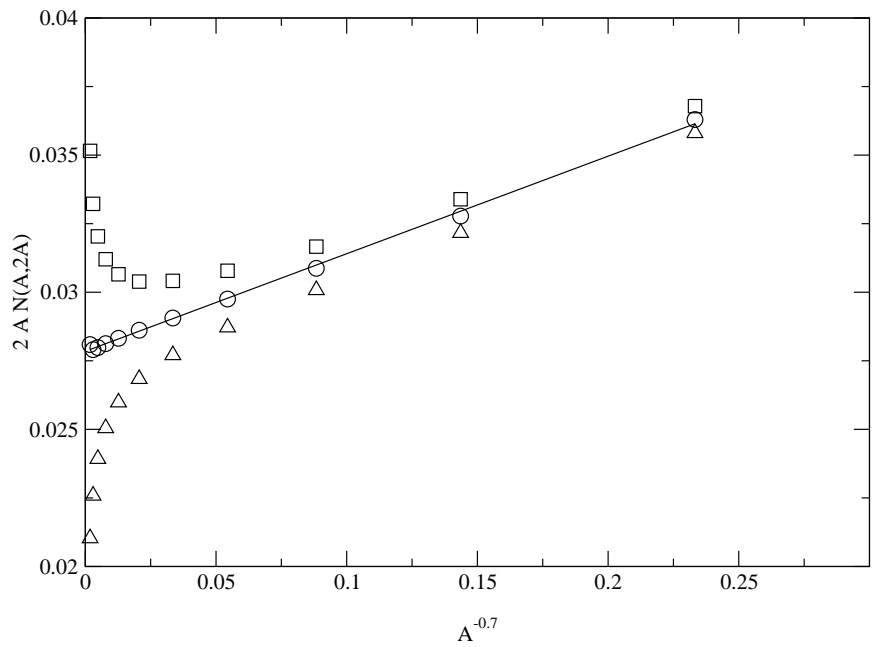}}
\caption{Plot of $2 A N(A,2A)$ vs.\ $A^{-0.7}$ for FK clusters of
the Q=3 Potts model:
external hulls (triangles), average (circles) and internal hulls (squares).}
\label{figFK3Potts}
\end{figure}

To thermalize the system, we used the Swendsen-Wang (SW) procedure of
identifying all FK clusters on the lattice and then randomly
reassigning their spins.  Indeed, the FK hull measurements and the
SW update method naturally go  hand in had in this
calculation, since the identification of the FK clusters
is needed for the SW method. For $Q=2$
and $Q=3$ we used a lattice of size $512 \times 512$ and obtained the
results shown in Figs.\ \ref{figFKising} and \ref{figFK3Potts}, where
once again we find large discrepancies between internal and external
clusters, and take the average of the two.  That average is
found to fall on a nearly straight line when plotted as a function
of $A^{-\theta}$ taking $\theta = 0.875$ for $Q=2$ and $\theta = 0.7$
for $Q=3$.  The extrapolated exponents
 are seen to approach the expected theoretical values, as 
shown in
Table \ref{tableC}.  We simulated $82,000$
samples ($Q=2$) and 1,000,000 samples ($Q = 3$).

Note that the discrepancy between internal and external hulls
reflects an inherent asymmetry for FK clusters of the Potts model
in finite periodic systems for $Q > 1$.
This asymmetry is also manifested in the behavior of the fraction
of bonds that are occupied, which in finite systems has a value somewhat
greater than the infinite-system value of
$\frac12$ \cite{HuChenIzmailianKleban}.

\begin{figure}
\centerline{
\epsfxsize=4.5in 
\epsfbox{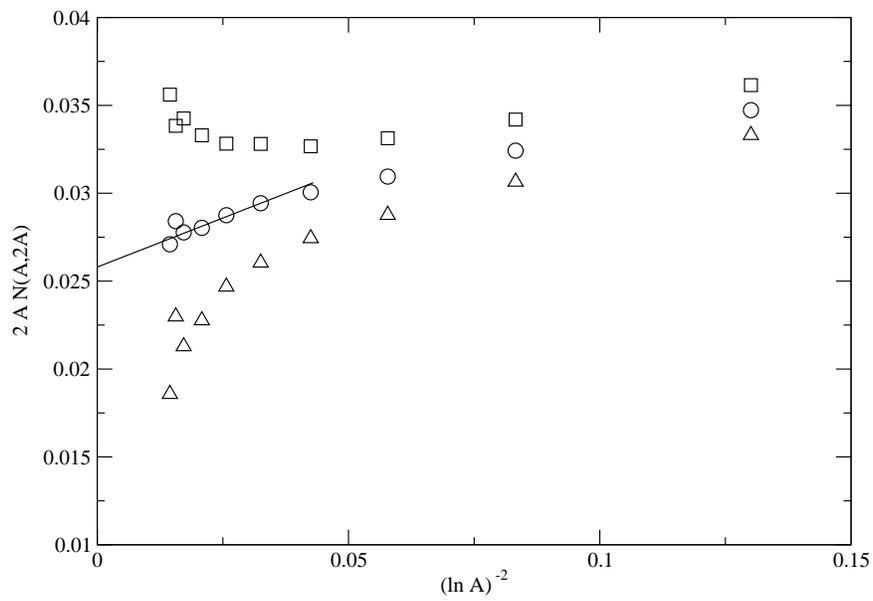}}
\caption{Plot of $2 A N(A,2A)$ vs.\ $(\ln A)^{-2}$ for FK clusters of
the Q=4 Potts model:
external hulls (triangles), average (circles) and internal hulls (squares).}
\label{figFK4Potts}
\end{figure}

For $Q = 4$, very large differences between internal and external 
hulls persisted even for relatively small values of $A$
on the $512 \times 512$ lattice, so we went to
 a larger lattice of size $2048 \times 2048$
(20,000 realizations) which improved the behavior
somewhat.  Even for this lattice, however,
large finite-size effects were apparent.
Similar large finite-size corrections have been seen in other Potts model
studies at $Q = 4$ (e.g., \cite{HuChenIzmailianKleban,Arguin02})
and are generally expected to be logarithmic in character.
In the Appendix we have calculated these corrections analytically
for this case and find 

\begin{equation}
N(A)\sim {C\over A}\left(1-{2a_2\over(\ln A)^2}+O((\ln A)^{-3})\right)
\label{logcorr}
\end{equation}
where $a_2$ is a constant. The above result implies that $2 A N(A,2A) =
C + O((\ln A)^{-2})$.
In Fig.\ \ref{figFK4Potts} we plot our results for
$2 A N(A,2A)$ as a function
of $(\ln A)^{-2}$.  The data fall on a straight line for
large $A$, and the intercept yields $C = 0.0258$, which is comparable
to our predicted value of $1/4 \pi^2 = 0.0253\ldots$.  

Note that if we  plot the data versus $1/\ln A$, we find
about as good of a fit to linear behavior for large $A$, but then the intercept
would $0.0231$, quite a bit below the predicted value of $C$.  
Likewise, if we fit the data to a power-law as we did for
other values of $Q$, we find fairly linear behavior with an
abcissa of $A^{-0.5}$, but now the intercept is $0.0279$.  Thus, 
the data is consistent with our prediction for $C$ combined
with the predicted $1/(\ln A)^2$ finite-size scaling.

\section{Conclusions.}

We derived and numerically confirmed predictions for the behavior
of the area-size distribution of various Potts model including
percolation clusters. 
For the latter, we also considered different lattices and percolation
types (site and bond) to demonstrate universality.
  The theoretical ideas presented in Section \ref{theory}  were well verified 
numerically,
especially in the percolation and Ising model cases. 
For $Q = 4$, our results were consistent with the logarithmic
finite-size behavior predicted here.

This work confirms the idea of a universal size distribution
expressed by  Eq.\ (\ref{Narea}). 
An alternate way to state that result is as follows:
Consider that the unit of area is now some value $A$ much smaller
than the lattice size (which is therefore no longer of unit area).  
Then,  (\ref{Narea}) implies that the number of clusters whose
enclosed area is greater than $A$, per unit area $A$, 
is a
constant $C$, for all values of $A$.  
The lack of dependence on $A$ is a direct
consequence of the scale-free nature of this fractal system.

The arguments put forward in Sec.~\ref{depth} also imply that the number
of cluster hulls which must be crossed to connect a typical point
deep inside the system to the boundary behaves as $4C\ln(L/a)$, where
$L$ is the system size, with same value of $C$ for each universality
class. So, for example, the fact that $C$ for critical Ising spin
clusters is half that for percolation clusters means, according to
Zipf's law, that the $n$th largest cluster is the Ising case has roughly 
half the area of the $n$th largest percolation cluster. This is
consistent with the fact that we have to cross one half as many cluster
hulls to reach the boundary in the Ising case. It might suggest that we
may go from the ensemble of percolation hulls to those of Ising
clusters simply by erasing every other percolation hull, e.g.~by ignoring
all the internal hulls! This however is not the case, as percolation
hulls have a different fractal dimension from those of Ising clusters.

The form of (\ref{Narea}) is also consistent with the existence of the 
universal amplitude ratio,
$R_\xi^+  = [\alpha(1-\alpha)(2-\alpha) {\cal F}_c]^{1/d} \xi_0$,
where $\alpha$ is the free-energy critical
exponent,
${\cal F}_c$ is the critical part of the free energy per unit area, 
and $\xi_0$
is the amplitude for the correlation length \cite{PrivmanHohenbergAharony}.
For any value of $Q$ in the random cluster model,
$\partial{\cal F}/\partial Q$ gives the mean total number of clusters
per unit area  $\sum_s n_s = \sum_A N_A$. 
At the critical point, $N_A\sim-N'(A)\sim C/A^2$ for $A\gg a^2$,
and near the critical point one expects a scaling law 
$N_A=A^{-2}\Phi(A/\xi^2)$, where $\Phi(u)$ is some nontrivial
scaling function with $\Phi(0)=C$, which decays
exponentially fast as $u\to\infty$. 
This gives, on substitution into $\sum_A N_A\sim\int_{a^2}^\infty
N_AdA$, 
\begin{equation}
\label{hyper}
\sum_s n_s\sim {\rm const.\ } + B\xi^{-2}
\end{equation}
where  the constant is nonuniversal, as it depends on the details of the
cutoff, and
\begin{equation}
B=\int_0^\infty{\Phi(u)-C\over u^2}\,du\,.
\end{equation}
Eq.\ (\ref{hyper}) is of the form expected from hyperscaling
 \cite{PrivmanHohenbergAharony}, with
$B = (R_\xi^+)^2/]\alpha(1-\alpha)(2-\alpha)]$ directly related to the
universal combination $R_\xi^+$ (recently given exactly for percolation
by Seaton \cite{Seaton}). \  However, we see that it is given by a
certain integral over a nontrivial scaling function, while $C$ is just 
one limiting value of this function. 

The results presented here represent the first examples where a measure of the 
cluster size distribution is given exactly (in the asymptotic limit),
both in exponent (here, simply  $-1$) and amplitude (the value $C$).
The agreement between the theoretical prediction and the numerical
results for percolation
(to a relative accuracy of better than $10^{-4}$) compares well
with other precision tests 
of conformal field theory predictions for
percolation amplitudes, for example the crossing
formula \cite{JCcrossing} (where the results have been
confirmed within a relative error of about $10^{-3}$ \cite{Langlands94,Ziff96}).
Knowing the exact result  for $C$
at the critical point allows finite-size
effects and behavior away from  the critical point to be studied,
without at the same time having to determine these critical
parameters.  In percolation especially there has been great interest
in size distributions and their finite size corrections, so this result
should be useful in that field.

\section{Acknowledgements}
J. C. thanks H. Kesten for useful comments and for drawing attention to
Refs. \cite{ChayesChayesDurrett,KestenZhang}. He was supported
in part by EPSRC Grant GR/J78327.
R.~Z. acknowledges a discussion
with Greg Huber concerning the origin of the finite-size corrections.

\section*{Appendix. Logarithmic corrections for $Q=4$.}

We summarize the arguments leading to Eq.~(\ref{logcorr}).
It has long been known that many critical quantities in the 4-state 
Potts model exhibit confluent logarithmic corrections. In the RG 
framework, this is explained by the existence of a marginally
irrelevant scaling variable \cite{wayback}. A general formalism for
computing the form of these corrections was developed in Ref.~\cite{CNS},
was taken further in Ref.~\cite{SalasSokal},
and recently has been applied to the fractal properties of $Q=4$ FK
clusters by Aharony and Asikainen \cite{AA}. In general \cite{CNS}, 
logarithmic corrections to susceptibilities take the form of
multiplicative powers of logarithms, and are therefore numerically
very significant,
but in some quantities, for example the finite-size scaling of the free energy
at the critical point \cite{Cardylogs}, 
they give only additive corrections. We shall argue
that this is the case here.

Following \cite{Cardylogs}, suppose that the fixed-point hamiltonian
is deformed by a marginal perturbation
${\cal H}^*\to{\cal H}^*+g\sum_R\Phi(R)$, where $\Phi$ is a scaling
operator with scaling dimension $x_\Phi=2$. We may develop the
current-current correlation function (\ref{twopoint} in a power series in $g$,  
the coefficient of each term being a sum over the $R_j$ of 
correlation functions
$\langle J_\mu(r_1)J_\nu(r_2)\Phi(R_1)\rangle$, 
$\langle J_\mu(r_1)J_\nu(r_2) \Phi(R_1)\Phi(R_2)\rangle$, and so on, each 
evaluated with respect to the fixed point hamiltonian. 
The form of the $r_1$ and $r_2$-dependence of
each of these correlation functions is completely fixed
by conformal invariance in two dimensions, so that they may be computed
in a simple model. Choosing a gaussian theory with hamiltonian
${\cal H}^*=\frac12\int(\partial\phi)^2d^2\!r$, a conserved current
$J_\mu\sim\partial_\mu\phi$, and the marginal operator
$\Phi\sim (\partial\phi)^2$, all the
correlators may be evaluated using Wick's theorem. 
For the $O(g)$ correction, it turns out that
the only non-zero components (in complex coordinates) have $\mu=z$,
$\nu=\bar z$, and \em vice versa\em. The form of the correlation
function is
\begin{equation}
\label{JJPhi}
\langle J_z(z_1)J_{\bar z}(\bar z_2)\Phi(0)\rangle
\propto 1/(z_1^2{\bar z_2}^2)
\end{equation}
where we have set $R_1=0$ for convenience.

Now the $O(g)$ correction to the total area within all loops (\ref{intJJ}) is 
\begin{equation}
\label{Og}
g\int \langle J_y(z_1)J_y(\bar z_2)\Phi(R_1)\rangle|x_1-x_2|\delta(y_1-y_2)
dx_1dx_2dy_1dy_2d^2\!R_1
\end{equation}
where $J_y\propto J_z-J_{\bar z}$. This is to be evaluated in a large
but finite region of linear size $O(L)$. As before, we shall use the infinite
volume continuum limit form (\ref{JJPhi}) of the correlation function,
justifying this \em a posteriori\em. The integral in (\ref{Og}) is then
proportional to the area $\cal A$ of the system, and we remove this
factor by setting $R_1=0$.
The remaining integral is then proportional to 
\begin{equation}
\int_{-\infty}^\infty {|x_1-x_2|\over (x_1+iy_1)^2(x_2-iy_1)^2}\,dx_1dx_2dy_1
\end{equation}
The contour integration over $y_1$ vanishes unless $x_1$ and $x_2$ have
the same sign: the result is then proportional to 
\begin{equation}
\int_0^\infty\int_0^\infty {|x_1-x_2|\over (x_1-x_2)^3}\,dx_1dx_2
=\int {dx_1\over x_1}\sim \ln(L/a)
\end{equation}
with an equal contribution from $x_1,x_2<0$. 
We have cut off the logarithmically divergent integral in the
last step, arguing that because the divergence is only logarithmic, it
was permissible to use the infinite-volume forms for the correlation
function in the integrand.

The important point about this result is that it is $O(g\ln L)$, not
$O(g(\ln L)^2)$, as might have been expected (recall that the leading
term is $O(g^0\ln L)$). A similar, but more tedious, calculation shows
that the next term is $O(g^2\ln L)$, and we conjecture that the
$n$th order term is $O(g^n\ln L)$. This is consistent with the
fact that, in the gaussian model, $g$ is exactly marginal so that $k$,
the coefficient of the $O(\ln L)$ term,
depends continuously on $g$. 

However, in the 4-state Potts model the perturbation is not exactly
marginal, and instead flows logarithmically slowly to zero under the RG.
This may be taken into account \cite{Cardylogs} by
replacing the bare expansion parameter $g$ by the running coupling
\begin{equation}
\tilde g(L)={g\over 1+bg\ln L}\sim (b\ln L)^{-1}+O(g^{-1}(\ln L)^{-2})
\end{equation}
where $b$ is a known constant whose value is not important.

Inserting this result into the formula (\ref{intJJ}) for the total area
$\langle A_{\rm tot}\rangle\sim{\cal A}\sum_{A<O(L^2)}N(A)$ gives 
\begin{equation}
\sum_{A<O(L^2)}N(A)\sim 2C\ln L\left(1+{a_1\over\ln L}+
{a_2\over(\ln L)^2}+O((\ln L)^{-3})\right)
\end{equation}
where the $a_j$ are non-universal constants.
Differentiating this with respect to $L^2\sim A$ then gives the main result
quoted in (\ref{logcorr})
\begin{equation}
N(A)\sim {C\over A}\left(1-{2a_2\over(\ln A)^2}+O((\ln A)^{-3})\right)
\end{equation}
The interesting feature of this result is the absence of the
$O((\ln A)^{-1})$ term, proportional to $a_1$.

\end{document}